\documentclass[12pt,halfline,a4paper]{ouparticle}
\usepackage[round]{natbib}
\usepackage{hyperref}

\newenvironment{itemize*}%
  {\begin{itemize}%
    \setlength{\itemsep}{0pt}%
    \setlength{\parskip}{0pt}}%
  {\end{itemize}}

\newenvironment{enumerate*}%
  {\begin{enumerate}%
    \setlength{\itemsep}{0pt}%
    \setlength{\parskip}{0pt}}%
  {\end{enumerate}}

\newcommand{\ind}{\mathbf{1}}

\newtheorem{theorem}{Theorem}

\newtheorem{assumption}[theorem]{Assumption}

\begin{document}

\title{\Large \textsc{What You See is What You Get:\\ Local Labor Markets and Skill Acquisition}\footnote{I would like to thank Ed Glaeser and Elhanan Helpman for their feedback and support. I would also like to thank Gordon Hanson and Marc Melitz.}}

\author{Benjamin Niswonger\footnote{Niswonger: Department of Economics, Harvard University, 1805 Cambridge Street, Cambridge, MA 02138 (e-mail: niswonger@g.harvard.edu).}}

\abstract{This paper highlights the potential for negative dynamic consequences of recent trends towards the formation of ``skill-hubs". I first show evidence that skill acquisition is biased towards skills which are in demand in local labor markets. This fact along with large heterogeneity in outcomes by major and recent reductions in migration rates implies a significant potential for inefficient skill upgrading over time. To evaluate the impact of local bias in education in the context of standard models which focus on agglomeration effects, I develop a structural spatial model which includes educational investment. The model focuses on two sources of externalities: productivity through agglomeration and signaling. Both of these affect educational decisions tilting the balance of aggregate skill composition. Signaling externalities can provide a substantial wedge in the response to changes in skill demand and skill concentration with the potential for substantial welfare gains from a more equal distribution of skills.

\vspace{3mm}

 }

\date{\today}

\maketitle
\thispagestyle{empty}
\vfill

\pagebreak

\setcounter{page}{1}
\section{Introduction}
Alfred Marshall's statement that ``when an industry has thus chosen a locality for itself...the mysteries of the trade are no mysteries; but are as it were in the air..." (\cite{MarshallA2013Poe}) has become the jumping off point for modern thinking on agglomeration\footnote{See for instance \cite{EllisonGlenn2010WCIA}, \cite{DavisDonaldR2019ASKE}, \cite{KantorShawnEverett2014Ksfr}, \cite{KlinePatrick2014LEDA}, \cite{ZacchiaPaolo2020KStN}}. Recently, it has formed the basis for a barrage of papers that have been written on the potential for place based policies in response to recent trends of regional divergence and the concentration of skilled labor documented in \cite{BerryChristopherR2005Tdoh} and \cite{MorettiEnrico2013Rwi}. Policy prescriptions based on these research efforts vary widely from relatively laissez faire (\cite{GlaeserGottlieb2008}) to more interventionist (\cite{RossiHansbergEsteban2019}). The common question being whether the benefits of agglomeration stemming from attracting skilled workers to a particular area outweigh the costs including the agglomeration losses in the region of origin (i.e. brain drain). These papers typically incorporate a static spatial model focusing on the productive benefits of concentrating skilled workers. 

This paper builds off of this scaffolding in order to look more closely at the impact of local labor markets on skill acquisition. In this sense, I am addressing the rest of Alfred Marshall's influential proclamation: ``so great are the advantages which \textbf{people following the same skilled trade} get from near neighbourhood to one another... children learn many of them unconsciously". The implication here is that people seem to be naturally predisposed to follow in the footsteps of the trades that people around them are engaged in. There is a positive externality exerted when people are surrounded by in-demand skills which leads them to pursue an education with high expected returns. The dynamic impact of concentrating labor in skill hubs is that it increases the likelihood that people will acquire in demand skills in those select few labor markets at the expense of reducing the likelihood elsewhere. If the impact of local labor markets on skill acquisition is concave, then this will reduce the overall pool of in-demand skills in the economy. 

In order to encourage future research along this vein, this paper provides evidence of the importance of understanding the impact of local labor markets on skill choice and provides the first steps in analyzing its importance relative to the standard focus on agglomeration. I first highlight the disparate outcomes associated with different majors, which goes a long way towards explaining findings that students have a difficult time predicting expected incomes by major as documented in \cite{MATTHEWWISWALL2015DoCM}. In response to this uncertainty, students tend to be biased towards the majors and occupations that they are able to observe in their local labor markets. The goal here is to stress the fact that with the variation across majors being nearly as large as the overall college skill premium, bias towards particular majors/degrees can have a major impact on future earnings. Spatial divergence by major and occupation is also exacerbated by recent trends towards more limited migration. 

However, even if there is bias towards local labor market skills beyond the local wage, this may not be sufficient to call for a more even spatial distribution of skills. First, it must be the case that the impact is concave such that the benefits of concentrating skills are smaller than the benefits of more evenly distributing skills. Secondly, concentrating skills may still create a high enough wage premium to overcome local skill bias. In order to consider this tradeoff more carefully, I develop a structural spatial model with two key externalities. The first is the standard agglomeration externality by which an increasing fraction of skilled labor in a local labor market increases the productivity of local workers. Second, I introduce a new externality that comes from signaling the expected wage associated with particular skills across potential locations. Although I frame it in terms of wage signaling, the reduced form evidence I provide is isomorphic to a broad range of potential biases with the common feature that students tend to be more likely to get the skills demanded in their local labor market beyond features related to the wage. 

Rather than calibrate the full structural model, I make several clarifying assumptions to allow me to present some suggestive reduced form evidence of the importance of this channel. This analysis is not meant to end the discussion through precise estimates of biases due to local labor market conditions but merely to introduce it. In my simplified model using decennial Census data for the years 1980,1990 and 2000, I find evidence that both the agglomeration and signaling externalities are of similar magnitudes with the important feature that concentrating skills has a heterogeneous impact across locations through the signaling externality. The effect of a more even distribution of skills across locations is convex in the initial skill concentration across locations. MSAs with low skill levels before redistribution experience large positive gains relative to initially high skill MSAs which experience small losses. Overall, this suggests that policies which affect the distribution of skills across space should consider not just the impact on productivity but on how the local labor market will impact skill acquisition more broadly.

\textbf{Related Literature:} This paper speaks to three distinct strands in the literature. Most directly this relates to the literature on place based policies. As alluded to above, many papers have taken the presence of agglomeration as inspiration to assess whether the government should be involved in shifting where people live. Traditionally the justification for concentrating workers in a particular location has hinged on whether the benefits of agglomeration are convex (\cite{GlaeserGottlieb2008}). More recently, the focus has been on whether skilled workers provide greater value to other skilled workers than to unskilled workers. When defining skilled workers as those performing cognitive, non-routine (CNR) tasks, \cite{RossiHansbergEsteban2019} finds that the optimal policy involves creating cognitive hubs by subsidizing CNR workers to move to large skilled cities and taxing non-CNR workers in those cities. 

Similarly, \cite{Fajgelbaum2020} find that the optimal policy involves increasing the concentration of skilled workers however, they achieve this by having both college educated and non-college educated workers moving away from the largest city. In this way all cities end up with a higher share of skilled workers. My baselines findings are roughly in line with this result. Note that this is the opposite of recent trends in the United States as described in \cite{MorettiEnrico2012Tngo}. This paper intends to add additional motivation to this debate by considering the downstream ramifications of the recent concentration of skilled workers in cognitive hubs on the future supply of skilled workers through signaling externalities. 

This paper also connects to the vast literature on educational choice. The ground breaking paper by \cite{MATTHEWWISWALL2015DoCM} showed that students at NYU had inaccurate beliefs about expected wages for different majors. A major finding was that students adjust correctly when provided updated information, but that the majority of the variation in major choice was not explained by perceived wage or ability. Alternative factors which affect choice of major include exposure to role models (\cite{Porter2020}) and industry experience (\cite{Boudreau2019}). This is similar to the \cite{Bell2019} findings for innovation in which early exposure to inventors increased the likelihood of inventing in the corresponding field. One way to view my work is as an attempt to begin filling in the gap of the ``taste residual" of educational choice.

Lastly, I hope to bring some additional perspective on the divergence in wages between skilled and unskilled workers. \cite{Autor2019} summarizes a broad swath of recent research concerned with the future of work. It highlights the on-going trend towards higher skill premiums and higher skill share cities. The importance of acquiring in demand skills has grown dramatically as the middle class jobs that provide a substantial buffer between high and low income groups has been automated (\cite{Autor2006}). This growing inequality is exacerbated by the relatively small gains in educational attainment made in recent years as discussed in \cite{Autor2020}. This problem is further exacerbated by the large discrepancies in wages by major \cite{Long2015} and differences in responsiveness to changes in the wage. My findings suggest that local bias will tend to push towards regional divergence and a disconnect between the aggregate skill premium and skill supply. 
\label{sec:intro}

\section{Suggestive Evidence}
The importance of location \textbf{and} college major loom large in the economy. In this section, I hope to highlight the importance of location \textbf{on} major choice and how this has the potential to significantly distort labor market outcomes. These facts will be used as a starting point for the model that will be developed in Section \ref{sec:model}.

\subsection{Complexity in College Major Choice for Potential Income}
It is especially important to study the determinants of skill investment because it is typically a one shot game with significant impact on lifetime income. This means that the plethora of potential mistakes won't be ironed out over time through repeated efforts. The decision of whether or not to go to college has because an increasingly significant decision as the skill premium has risen as documented in \cite{Katz1992}. The decision is further complicated by the choice of major. The ratio of returns to high income vs. low income majors is on the same order as the college wage premium (\cite{Altonji2012}). Thus, the question of choosing the ``right" major can be just as important as the decision of whether or not to go to college, and all of these decisions are typically made once in a lifetime. 

\begin{figure}[h]
\centering
\caption{ACS Total Income by Major for Prime Aged College Graduates}
\label{fig:mean_wage_by_major}
\includegraphics[height = 3.5in]{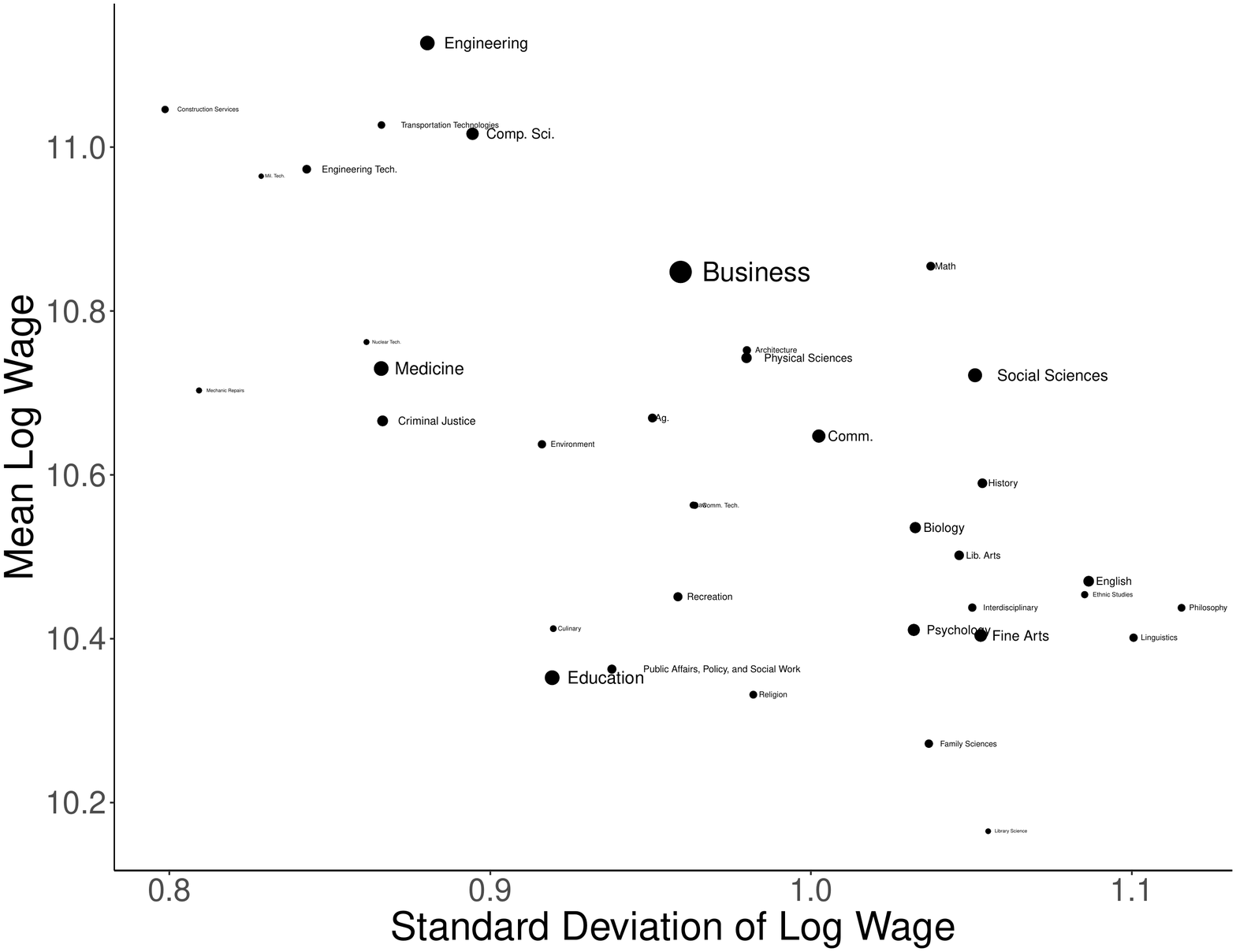}
\caption*{\footnotesize This figure shows the mean and standard deviation of log of total annual wage and salary for employed workers with a college degree between the ages of 25 and 55. The data comes from the ACS for the years 2010-2018 and were acquired through IPUMS.}
\end{figure}

This paper is not intended to do a deepdive on what drives the differences in returns (for more on this see \cite{Altonji2012} or \cite{Andrews2022}), but rather I focus on the spatial component of skill acquisition. However, it is useful to highlight the extreme differences in outcomes. Using data collected by IPUMS from the American Community Survey (ACS) for the years 2010-2018, I show in figure \ref{fig:mean_wage_by_major} the high level of variation of mean wage for workers from 25-55 (prime aged) with a college degree \textbf{across majors}. The highest earning major is engineering with an average income over this time period of roughly \$90K whereas the lowest earning major was library science with a mean total income of \$36K. There is also considerable heterogeneity in terms of the risk associated with the different majors. The standard deviation ranging from \$26K for library science to \$82K for social sciences. In figure \ref{fig:mean_wage_by_major} we can see the trade off in terms of the mean log wage and the standard deviation of the log wage. The majors range drastically both in terms of mean and standard deviation as well as in the tradeoff between these dimensions. This evidence is roughly in line with \cite{ChristiansenCharlotte2007Trti} which further argues that there is an efficient frontier in terms of the risk reward tradeoff for particular degrees. They take the fact that people choose degrees which are inside of the frontier as evidence for the importance of taste in educational attainment. 

For the purposes of this paper, what is clear is that making an efficient educational choice requires substantial information about future earnings especially when there are concerns about the risk of particular majors. \cite{MATTHEWWISWALL2015DoCM} finds that risk aversion is quantitatively important. This is also true at the occupational level as in \cite{DillonEleanorW2018RaRT} where workers are willing to accept lower earnings in order to avoid wage and employment risks. Especially given the fact that the expected outcomes for the average person in the population may be less important than signals about one's own potential outcome in evaluating a potential major, it is unsurprising that students have limited accuracy in terms of expected outcomes for the aggregate population. 

\subsection{College Major Choice is Consistent with Local Degree Distribution }

Given the large distribution of potential outcomes based on college major, it is important to understand the underlying reasoning behind major choice. For instance, \cite{MATTHEWWISWALL2015DoCM} find that students respond to pecuniary factors but that the majority of the choice seems to be explained by taste parameters. In this paper, I argue that a large component of this apparent taste may be driven by growing up in a labor market with concentrations in particular skills. 

I provide reduced form evidence on the relationship between skill acquisition and local skill concentration using data from the Higher Education Research Institute (HERI) for the years 2000-2008. Specifically, I focus on the primary major stated by respondents of the True Freshman Survey, which has the broadest coverage in the survey data. This survey has roughly 300000 respondents each year and has been administered by over 1900 institutions. The data I look at includes both the stated major as well as the zipcode of each student's home address. This allows me to map respondents to their home counties and link the data to the ACS. Unfortunately, the ACS only includes degree field starting in 2009, so that in order to back out the number of prime aged workers by degree for the year 2000 I calculate the fraction of individuals in a particular occupation by degree for the years 2009-2018 and use this relationship to develop a proxy for local labor market skills. This allows me to approximate the number of people with a particular degree in a county using Census data for the year 2000.  

\begin{figure}[h]
\centering
\caption{Comparing Working Population Major Composition vs. College Student Stated Major}
\label{fig3}
\includegraphics[height = 3.5in]{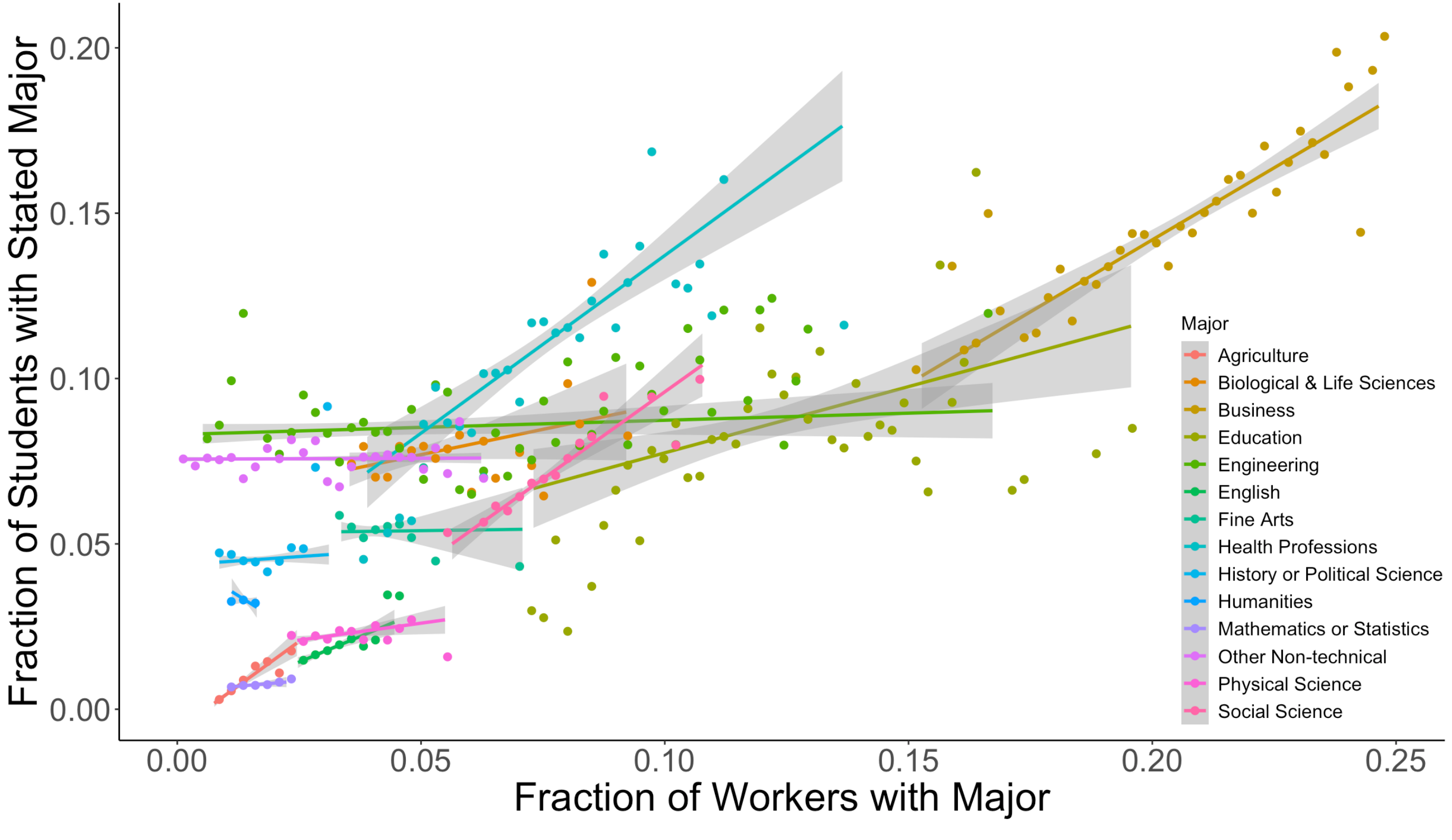}
\caption*{\footnotesize This figure plots the relationship between stated major in the TFS administered by the HERI and the fraction of workers in the students home county. The fraction of workers is approximated based on an occupational mapping from major to occupation from the ACS for years 2009-2018. The home county is based on a crosswalk from zipcode to county. In cases where there is a many-to-one match the student is mapped to each county. }
\end{figure}

I show that there is a clear and statistically significant positive relationship between the fraction of workers with a particular degree and the fraction of freshman who declare that as their major at the home county level. This relationship is plotted in Figure \ref{fig3}. Clearly, degree attainment is positively correlated with local labor conditions. This relationship is particularly strong in business, health, social science and education while being much flatter in STEM degrees. This is what one would expect given the high level of public attention on the returns to STEM degrees and the importance of STEM education. However, other degrees such as business, social science and medicine which are on the higher end of the major earnings distribution are less vocally promoted. This implies that exposure to occupations that requires those degrees are more likely to affect the decision to major in those fields. 

A simple regression of the fraction choosing a particular major on the fraction of workers with that major in their home county is shown in Table \ref{tab:maj_match}. The correlation between local labor market conditions and major choices is significant even after controlling for major and county fixed effects.

\begin{table}
\caption{Effect of Local Labor Market on Major Choice \label{tab:maj_match}}

\global\long\def\sym#1{\ifmmode^{#1}\else$^{#1}$\fi}%
 \begin{center}
\begin{tabular}{llll}

\hline 
 & \multicolumn{3}{c}{Fraction Choosing that Major}\tabularnewline
\hline 
Fraction of Workers with Major in County & 0.45\sym{{***}} & 0.16\sym{{***}} & 0.15\sym{{**}} \tabularnewline
 & (0.02) & (0.02) & (0.02)\tabularnewline
\hline 
Major FE & NO & YES & YES\tabularnewline
County FE & NO & NO & YES\tabularnewline
\hline 
Obs & 9395 & 9395 & 9395\tabularnewline
$R^{2}$ & 49.6\% & 66.1\% & 67.4\%\tabularnewline
\hline 
\multicolumn{3}{l}{{\tiny{}Robust standard errors clustered at the primary industry in
parentheses.}}\tabularnewline
\multicolumn{3}{l}{{\tiny{}\sym{*}$p<0.05$, \sym{**}$p<0.01$,\sym{***}$p<0.001$.}}\tabularnewline
\end{tabular}
\end{center}
\end{table}

Clearly, these results are partially explained by greater demand for particular occupations in specific home counties. Combined with the fact that it is costly to move between counties, this would be sufficient to drive the positive relationship documented in this section. However, such employment specific outcomes would seem to be especially important for technical occupations which require STEM degrees rather than in occupations that are more general purpose (business and social science) and services which are broadly needed (education and medicine). We see exactly the opposite trends in Figure \ref{fig3}. These results are in fact much more naturally explained by informational frictions. 

\subsection{Inter-County Migration has Slowed}

The local bias in education wouldn't be problematic if skills were redistributed across locations. This would reduce the local bias in skill composition. However, overall rates of inter-county migration have been slowing substantially in recent years as documented in \cite{Molloy2011}. There is some distinction however between skilled and unskilled migration. Using Current Population Survey data from the years 1985-2019, I show trends in the migration rates for people 25-55. I subset to this age range to approximate a time after receiving a degree and before retirement when migration ceases to have an impact on labor market composition.

\begin{figure}[h]
\centering
\caption{CPS Intercounty Migration Rates }
\label{fig:migration}
\includegraphics[height = 2in]{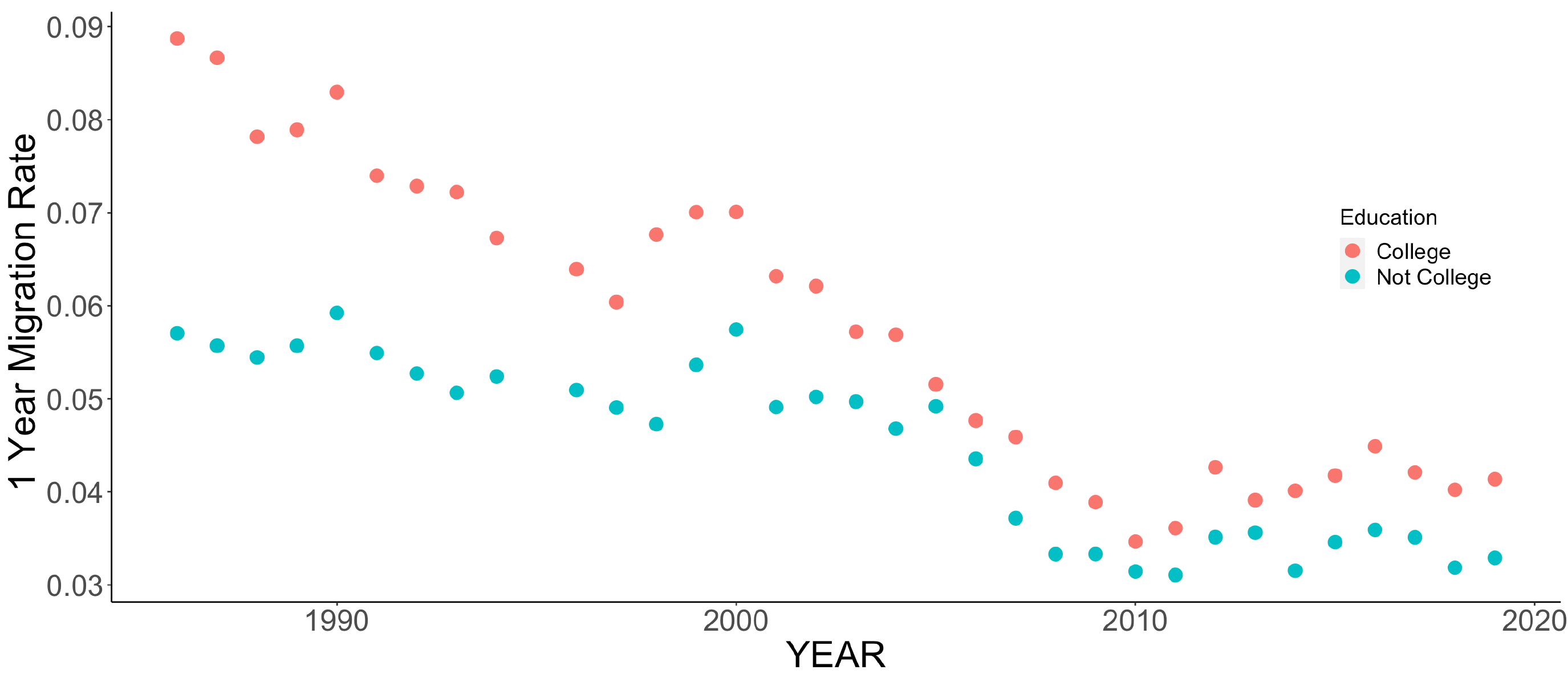}
\caption*{\footnotesize This figure is based on CPS data from 1985-2019. The population is subset to ages 25-55 with education levels based on whether or not the respondent had a college degree. }
\end{figure}

The downward trend shown in Figure \ref{fig:migration} is dramatic for both college and non-college educated workers. As of 2010, the United States seems to have reached a plateau of consistently lower migration rates although it is still the case that those with a college degree are more mobile. \cite{Stange2022} look at migration patterns for recent graduates more closely and find that according to publicly available LinkedIn data, 50\% of recent college grads stay in the MSA in which the institution they attended was located. This is especially true for less selective universities. The low migration rates post graduation are further exacerbated by the fact that the median 4-year public university student travels 18 miles to attend their institution according to the American Council on Education (\cite{Hillman2016}).

In summary, I have shown that there exists a large variation in the returns to skill both at the macro level (college vs. not) and at the micro level (college major). This variance explains findings that students do a poor job of predicting expected returns to specific majors. In the face of this uncertainty there seems to be a tendency to pick a major which is well represented in a students home labor market. The pile-on effect here is then exacerbated by decreased migration rates. In the next section, I develop a structural spatial model which captures these facts in order to show how local bias in skill acquisition can affect the geographic distribution of skills leading to persistent regional inequality. 
\label{sec:evidence}

\section{Structural Spatial Model} 
\label{sec:model}
The objective of this section is to build off canonical structural spatial models while including the impact of educational choice allowing for a positive externality associated with local labor markets. We will refer to the agents in this economy as students because the key decisions take place before completing their studies. Students choose type and level of education, \(m \in \{0,1,...,M\}\), where there are \(M\) majors and \(m=0\) refers to forgoing a college degree. Attaining a particular education results in an ability level, \(\alpha_{c_om}\), which depends on the location of origin, \(c_0\) and educational choice. Each student can choose to move to any city, \(c \in C\). In order to move from origin city \(c_0\) to destination \(c\), a moving cost of \(d_{c_0c}\) must be paid where \(d_{c_0c_0}=0\). Once in that location they inelastically supply labor and receive the prevailing wage which is a function of their ability. 

\subsection{Preferences:}

As in \cite{DavisDonaldR2019ASKE}, each student will consume \(\lambda\) units of a local non-tradeable good, a tradeable good and one unit of housing. They also have idiosyncratic tastes over major-city pairs denoted by \(z_{c_0mc}\), which can in principle depend on the origin. For tractability, I assume that these shocks are distributed independently Frechet across major-destination pairs such that
\[F_{c_0mc}(z) = e^{-T_{c_0mc}z^{-\theta}}\]
Absolute preference for particular major-destination pairs varies by home location \(c_0\) through \(T_{c_0mc}\), and the dispersion of preferences is inversely related to \(\theta\). An important assumption for this analysis is that the idiosyncratic component of preferences is specific to educated workers and is realized after the decision of whether or not to get a college degree. This captures the fact that shocks which affect preferences and costs take place during university before having committed to particular majors or destinations. Although this is a significant assumption, it greatly simplifies the estimation procedure in Section \ref{sec:calibration}. This assumption is summarized in the following assumption: 

\begin{assumption}
The idiosyncratic taste component of preferences is only significant for skilled workers. 
\end{assumption}

In order to account for the risk that students face in deciding on what skills to invest in, students have CARA utility functions. This formulation allows for tractable results with the Frechet preference shock. Combined, this implies that student's indirect utility is given by
\begin{equation}
\label{eq:indirect_utility}
V_i(c_0,m,c) =-\frac{1}{z_{c_0mc,i}}\exp\left(- w_i(\alpha_{c_0m},c) + t_{c_0m}+ \lambda p_{nc} + p_{hc}-a_c +d_{c_0c}\right)
\end{equation}
which depends on the prevailing wage for workers with ability \(\alpha_{c_0m}\) in city \(c\), \(w(\alpha_{c_0m}c)\), as well as the price of nontradeables, \(p_{nc}\), housing, \(p_{hc}\), a local amenity \(a_c\) and the cost of education,\(t_{c_0m}\).

\subsection{Learning:}

This section introduces the main departure of the model from canonical formulations. There are many reasons why we might expect students to be affected by their local labor market. There may be direct effects along the lines of Marshall's quote in the introduction: being around people with a particular education might endow you with the necessary ability to succeed. This would effectively reduce the cost of obtaining particular skills. It may also be the case that people surrounded by certain occupations grow to like them better. This would add an amenity term for particular occupations. While both of these are reasonable explanations, I will focus on a third which is that being surrounded by people with particular skills reduces the uncertainty about obtaining those skills. 

The reduction in uncertainty is in line with findings in the education literature which highlights the lack of information students have about future earnings and their risk aversion (\cite{Altonji2016}, \cite{Andrews2022}, \cite{Altonji2012}). Specifically, \cite{MATTHEWWISWALL2015DoCM} find that students exhibit a high level of risk aversion which significantly affects their major choice. The source of uncertainty that is revealed by local labor markets may be with regards to the returns to ability or to the cost of acquiring skills. This model will be framed in terms of uncertainty about the wage although it is isomorphic to alternative interpretations. The general form that this derivation simplifies to is both a gift and a curse in that one needs to either be agnostic about the underlying reason for the effect or must lean on the identification strategy in order to pick out a specific component. For more discussion about the learning channel and some reduced form evidence see Appendix Section \ref{sec:learning}.

\subsection{Signal Extraction Model:}
From the standpoint of the individual, the wage is taken as given although there is uncertainty as to what the student's future wage will be when they enter the labor market. I assume that students consider the wage to be a normal random variable which depends on your ability (origin-major choice) and the destination specific production technology. The wage they are trying to learn about takes the form: 

\[w(c_0,m,c) = \mu_{\alpha_{c_0m}c} + \xi_{c_0mc}\]
where \(\mu_{\alpha_{c_0m},c}\) is the mean wage and \(\boldsymbol{\xi}\) is the variance which follows a multi-variate normal distribution, \(\boldsymbol{\xi} \sim N(0,\boldsymbol{\Sigma})\). The variance for a given origin-major-destination wage is given by \(\sigma_{c_0mc}^2\). The common prior about the distribution of mean wages is given by \(\mathbf{w_0} \sim N(\mathbf{\mu_0} ,\boldsymbol{\Sigma})\). Here \(\mathbf{w_0}\) is a \(C\text{ x }C\text{ x }M\) vector of wages, which includes all possible combinations of majors across origin and destination cities. When making their education/location decision an individual from \(c_0\) will only use information about the relevant \(1\text{ x }C\text{ x }M\) components of the wage vector. Similarly the correlation structure of wages across locations is given by \(\boldsymbol{\Sigma}\), which is known.

Learning takes place by observing a discrete number, \(\kappa\), of observations drawn randomly from the origin city. I'll assume that there is migration across destinations such that the wage observed depends on migration patterns: if more people from Boston move to San Francisco, people in Boston become more aware of what wage they would be expected to receive in San Francisco. Specifically, let \(\boldsymbol{M}\) be the migration matrix which defines the fraction of individuals from the previous generation that moved across cities where \(M_{cc'}\) is the fraction of individuals who moved from \(c\) to \(c'\). The vector of the number of observations that someone from origin city \(c_0\) observes across destinations is given by \(\kappa \boldsymbol{M_{c_0}}\). We can define the fraction of individuals from \(c_0\) with major \(m\) to be \(\delta_{c_0m}\). For each location they observe \(\kappa M_{c_0c}\delta_{c_0m}\) observations of the origin-major-destination specific wage signal:

\[w(c_0,m,c) = \mu_{\alpha_{c_0m}c} + \xi_{c_0mc} + \hat{\xi}_{c_0mc}\]
where \(\hat{\xi}_{0mc}\) is the signal noise distributed \(N(0,\sigma^2_{\hat{\xi},c_0mc})\). The diagonal of the signal precision matrix is given by 
\[P_{c_0mc} = \left(\sigma^2_{c_0mc}+\sigma^2_{\hat{\xi},c_0mc}\right)^{-1}\kappa M_{c_0c} \delta_{c_0m}\]
Updating results in the following:
 \[\boldsymbol{w} \sim N\left((\boldsymbol{\Sigma}^{-1}+\boldsymbol{P})^{-1}(\boldsymbol{\Sigma}^{-1}\boldsymbol{\mu_0}+\boldsymbol{P}\boldsymbol{\mu})   ,(\boldsymbol{\Sigma}^{-1}+\boldsymbol{P})^{-1}\right)\]
The vector of mean wage signals is given by \(\boldsymbol{\mu}\). We'll denote the posterior distribution to be given by \(N(\boldsymbol{\mu^p}, \boldsymbol{\Sigma^p})\)
\subsection{Expected Indirect Utility:}
Now that we have put structure on the uncertainty faced by students in this model, we can evaluate expected indirect utility. In this model, the impact of local labor markets on education, beyond direct effects captured by the wage, come through agents' risk aversion. An individual will penalize the wage as a function of the number of signals of each major-destination combination that they observe. Taking expectations of Equation \ref{eq:indirect_utility} conditional on the Frechet shock results in:

\begin{align}E_{c_0}[V_i(c_0,m,c)] = \\ \nonumber -\frac{1}{z_{c_0mc,i}}\exp\left(-\mu^p_{c_0mc}+ \frac{1}{2}\left(\sigma^2_{c_0mc}+\Sigma^{p}_{c_omc}\right) + t_{c_0m}+ \lambda p_{n,c} + p_{h,c}-a_c +d_{c_0c}\right) \end{align}

We now assume that students have diffuse priors about the wage such that their only source of information is through wage signals that they receive. 

\begin{assumption}: 
Students have diffuse priors such that their posterior beliefs are entirely dependent on the signalling structure of their local labor market. 
\end{assumption}

Then we have that
\begin{equation}
\label{eq:posterior_cov}
\Sigma^{p}_{c_omc} = \frac{\left(\sigma^2_{\xi,c_0mc}+\sigma^2_{\hat{\xi},c_0mc}\right)}{\kappa M_{c_0c}\delta_m}\end{equation}
This makes clear that the indirect utility associated with a particular origin-major-destination is an increasing function of the number of workers in their local labor market with the relevant major and the level of migration to a particular destination. 

\subsection{Goods Production}

I abstract from the production side of the economy to consider only labor demand. We have already seen that wages are given by 
\[w(c_0,m,c) = \mu_{\alpha_{c_0m}c} + \xi_{c_0mc}\]
In this section I bring in the traditional productivity effects of agglomeration. There is a strict difference between \(m=0\) and all other major choices. Specifically, it is assumed that students who choose to forego a college degree will be employed in the non-tradable sector. Productivity in the non-tradable sector is the same across all origins and destinations. This sector can be thought of as the low-skill services sector. Because productivity is the same everywhere, it is easier to learn and thus, I'll assume that the wage for this sector is known with certainty. For all other workers, productivity takes the form of a destination specific productivity, \(\rho_c\), a normally distributed productivity draw \(h_{\alpha_{c_0m}c}\) and an agglomeration component taken to be the average of all individual productivity levels in the city. This is a simplified version of the framework in \cite{DavisDonaldR2019ASKE}. Combined, the mean wage for college graduates is given by 

\[w_{c_0mc} = \rho_c h_{\alpha_{c_0m}c} H_{cm}^{\gamma_{\alpha}} \]
\[H_{cm} = \frac{1}{L_c}\int_{\{i |c_i = c,m_i = m \}} h(\alpha_i,c) di \]
where for individual \(i\) from origin \(c_0\) with major \(m\), \(h(\alpha_i,c) = h_{\alpha_{c_0m}c}\) and \(L_c\) is the total population in destination \(c\).  This formulation highlights that students are learning about the overall city productivity vector \(\boldsymbol{\rho}\), the level of agglomeration \(\boldsymbol{H}\), and an origin-major specific wage component, \(\boldsymbol{h}\). This last piece can be thought of as a matching function between the specific abilities produced by a certain origin-major pair and the skills demanded across cities.

Altogether we have that the mean wage is given by 
\begin{equation}
w_{c_0mc}=\left\{\begin{array}{ll}
						  \rho_c h_{\alpha_{c_0m}c} H_{cm}^{\gamma_{\alpha}} & \text { if } m\neq 0 \\
						  p_{n,c} & \text { if } m = 0 
\end{array}\right.
\end{equation}

A key feature of this model will be how the productivity effects of agglomeration which come about by concentrating skills interacts with the signaling effect. 

\subsection{Housing}

The price of housing is given by a constant elasticity price equation which depends on the population in the city, \(L_c\). 

\begin{equation} 
p_{h,c} = \kappa L_c^{\gamma_h}
\end{equation}

\subsection{Solving the Model}

The solution to this model is a Nash Equilibrium, where all students make the optimal major-destination choice accurately anticipating the wage and labor markets clear. In order to simplify notation, I'll denote the effective wage \(\omega_{c_0mc}\) as the value of the wage taking into account uncertainty net of education cost:

\begin{equation}
\label{eq:effective_wage}
\omega_{c_0mc}  \equiv  \ind_{m>0}\left(\mu^p_{c_0mc}- \frac{1}{2}\left(\sigma^2_{\xi}+ \Sigma^p_{c_0mc}\right) - t_{c_0m}\right)+ \ind_{m=0}\left(p_{nc}\right)\end{equation}
This allows us to write the major-destination optimization problem as: 

\[\max_{m\in \{0,M\},c\in C} E_{c_0}[V_i(c_0,m,c)] = \max_{m\in \{0,M\},c\in C} -\frac{1}{z_{c_0mc}(i)}\exp\left(-\omega_{c_0mc}+ \lambda p_{n,c} + p_{h,c}-a_c +d_{c_0c}\right)\]

Taking advantage of the properties of the Frechet distribution, we can then calculate the probability that a student with a college degree from \(c_0\) chooses major \(m\) and city \(c\) as  

\begin{equation}
\label{eq:probability_of_location}
P(c,m|c_0) = \frac{T_{c_0mc}\exp\left[\theta\left(\omega_{c_0mc}-\lambda p_{n,c} - p_{h,c}+a_c -d_{c_0c}\right)\right]}{\sum_{m'\in \{0,M\}}\sum_{c' \in C} T_{c_0m'c'}\exp\left[\theta\left(\omega_{c_0m'c'}-\lambda p_{n,c'} - p_{h,c'}+a_c' -d_{c_0c'}\right)\right]}\end{equation}
For unskilled labor it must be the case that indirect utility is equalized across locations \(c\) for someone from \(c_0\). This implies 

\begin{equation}
\label{eq:unskilled_balance}
(1-\lambda)p_{nc}+a_c - p_{h,c} -d_{c_0c} = \overline{v}_{c_0} \quad \forall \quad  c\in C
\end{equation}
Because all students are ex ante identical we also have that there must be an indifference condition between getting a college degree and not. This condition is analogous to the price index in \cite{Eaton2002} so that the value of college is proportional to the ``market potential" with the appropriate elasticity.

\begin{equation}
    \label{eq:college_choice_eq}
    \overline{v}_{c_0} = \Xi \Phi_{c_0}^{1/\theta}
\end{equation}
where 
\[\Phi_{c_0}=\sum_{m'\in \{0,M\}}\sum_{c' \in C} T_{c_0m'c'}\exp\left[\theta\left(\omega_{c_0m'c'}-\lambda p_{n,c'} - p_{h,c'}+a_c' -d_{c_0c'}\right)\right]\]
From market clearing we have that 
\begin{equation}
L = \sum_c L_c
\end{equation}
and 
\begin{equation}
\delta_{cm}L_c =  \sum_{k\in C} \left[L_k P(c,m|k)\right]
+ \underline{L}_{cm}\end{equation}
where \(\underline{L}_{cm}\) is the initial number of workers in city \(c\) with major \(m\). This equation simply states that the number of people who end up with major \(m\) in city \(c\) must be equal to the number who decide to get that major and move to that city across all other location plus the initial endowment. 

Additionally, the market for non-tradable production must clear such that
\[\delta_{cm=0}L_c = \lambda L_c \quad \forall c \]
This condition will pin down the price of the non-tradable good and therefore, the low skilled wage across locations. 

\subsection{Comparing to the Existing Literature}
\label{sec:comparing_to_lit}
In order to compare this to previous work on agglomeration, I can simplify the model to focus on the choice of whether or not to obtain a college degree, \(m \in \{0,1\}\). The key point of emphasis is on the determinants of the effective wage, \(\omega_{c_0mc}\). We can rewrite Equation \ref{eq:effective_wage} substituting in for the determinants of the true wage in a setting where Assumption 1 holds:

\begin{equation}\omega_{c_0mc}  \equiv  \ind_{m=1}\left(\rho_c h_{c_0c}\delta_c^{\gamma_{\alpha}}- \frac{1}{2}\left(\sigma^2_{\xi}+  \frac{\left(\sigma^2_{\xi,c_0c}+\sigma^2_{\hat{\xi},c_0c}\right)}{\kappa M_{c_0c}\delta_{c_0}}\right) - t_{c_0}\right)+ \ind_{m=0}\left(p_{n,c}\right)\end{equation}
Note that I have used the fact that it is without loss to consider \(H_c = \delta_c\). Now we can see that the effective wage for college educated workers depends on the concentration of skill in the \textbf{destination} through a positive productivity effect, but it also depends on the concentration of skill in the \textbf{origin} through a signaling effect. The question of place-based policy from the perspective of optimal skill allocation then hinges crucially on the relative value of these two effects.

\section{Model Simplification and Estimation}
\label{sec:calibration}
The calibration I present in this paper is meant to be a first pass in understanding the impact of labor market signaling on skill acquisition. I leverage migration data from the ACS as well as local wages and housing prices in order to back the key structural parameters in the model. This will allow me to perform a counterfactual where we begin from an equal distribution of skills across locations. In order for the calibration to be consistent with recent work such as \cite{DiamondRebecca2016Tdaw}, I focus on the two skill case as in Section \ref{sec:comparing_to_lit}.

\subsection{Data}
\label{sec:data}
The primary moment that I will target in this analysis is given by Equation \ref{eq:probability_of_location} and represents migration shares across origin-destination pairs. With this in mind, the primary data used in this calibration is decennial census data for the years 1980, 1990 and 2000 accessed through IPUMS. In these years, a 5\% sample of the population was asked to state where they lived 5 years ago. This allows me to back out the migration matrix which forms the back bone of the calibration. 

I also use the education data from the census, which is limited to coarse measures of educational attainment and does not distinguish by major. Although I could proxy for choice of major by industry and occupation, this would lead to substantial noise. The ACS has data on major choice but has more limited migration information and is only available from 2009 onwards. This data also includes household rental rate which is used as a proxy for local housing prices. I also deflate all wages and prices using national consumer price index. 

\subsection{Non-Tradable Component and Amenities}

The first step in the calibration is to solve for \(\lambda\) which represents the amount of non-tradable consumed. The model suggests there are two distinct ways to calculate this. The first method is to realize that the indifference condition shown in Equation \ref{eq:unskilled_balance} implies a specific relationship between the local unskilled wage, \(p_{nc}\) and the price of housing, \(p_{hc}\), which depends on \(\lambda\). In order to simplify this analysis I'll make the additional assumption that the cost of moving is equal across destinations. Clearly, this will limit the models ability to capture the variation in migration patterns. However, this aides the step-by-step estimation of the model and will be validated in part by looking at the city-specific amenity levels that are dependent on this assumption. 

\begin{assumption}
There are no migration costs. 
\end{assumption}

This assumption leads to a simplified version of Equation \ref{eq:unskilled_balance}:

\begin{equation}
\label{eq:simplified_unskilled_balance} 
(1-\lambda)p_{nc}+a_c - p_{h,c} = \overline{v} \quad \forall \quad  c\in C
\end{equation}

Rearranging, this result yields an econometric means of evaluating \(\lambda\) by regressing the price of housing on the price of non-tradables which through the lens of the model is the unskilled wage. Using the Census data described in Section \ref{sec:data}, I estimate the following specification: 

\begin{equation}
    \label{eq:econometric_lambda}
p_{hct} = (1-\lambda)p_{nct}+a_c+\epsilon_{ct}
\end{equation}

Note that I've include a time subscript \(t\) to emphasize that observations are at the MSA-Year level. The value of \(p_{nc}\) is taken to be the average wage of workers in an MSA that do not have a college degree. Note that amenity term \(a_c\) is a county level fixed effect. I assume that \(\lambda\) and \(a_c\) do not change across the time period. The result of this analysis is a \(\lambda\) of 0.703. The regression also produces amenity values. In Table \ref{tab:amenities}, I list the top and bottom 5 MSAs in terms of their amenity value that comes out of the regression in Equation \ref{eq:econometric_lambda}. These MSAs provide some reassurance as some of the warmer and nicer areas in the country appear in the top 5 while some of the colder and poorer areas for this time period appear in the bottom 5.

\begin{table}[ht]
\caption{High and Low Amenity MSAs}
\centering
\begin{tabular}{ll}
\hline 
\multicolumn{2}{c}{Top 5}\\
    MSA & Amenity Value \\
  \hline 
  Santa Barbara, CA &  \$4560\\ 
  Washington D.C. &  \$4310\\ 
  Gainseville, FL &  \$4100\\ 
  San Jose, CA &  \$4010\\ 
  State College, PA &  \$3890\\ 
   \hline
\end{tabular}
\quad
\centering
\begin{tabular}{ll}
  \hline \multicolumn{2}{c}{Bottom 5}\\
  MSA & Ameinty Value \\
  \hline 
  Flint, Michigan & -\$3170 \\ 
East Chicago, IN & -\$2520 \\ 
Detroit, MI & -\$2060 \\ 
Sheboygan, WI & -\$1710 \\ 
Green Bay, WI & -\$1670 \\ 
   \hline
\end{tabular}
\caption*{\footnotesize Results are based on a regression of of annualized rents on low skilled wages across MSAs from Census data for the years 1980, 1990 and 2000. Amenity values are on anualized basis. }
\label{tab:amenities}
\end{table}

An alternative means of calculating \(\lambda\), is noting that in aggregate it must be that \(\lambda = \frac{L_{m=0}}{L}\). With this alternative formulation I get \(\lambda = .729\). The list of high and low amenity MSAs is roughly the same with Honolulu, HI replacing State College, PA in the top 5. These differences do not fundamentally affect the analysis and going forward I will lean on the first specification with \(\lambda = .703\). It is important to note that both of these estimates and the subsequent amenity values are heavily dependent on the choices made in our structural model. Future work may perform a more robust analysis as in \cite{DiamondRebecca2016Tdaw}. However, it is beyond the scope of this analysis. 

\subsection{Signaling Impact}

I can evaluate the impact of the fraction of college worker's in a students local labor market on their educational choice by simplifying and estimating Equation \ref{eq:probability_of_location}. To simplify this equation it is useful to take logs and simplify for the binary education decision resulting in

\[\ln P(c|c_0) = \ln T_{c_0c}+\theta\left(\omega_{c_0c}-\lambda p_{nc} - p_{hc}+a_c -d_{c_0c}\right) - \ln(\Phi_{c_0})\]
where \(\Phi_{c_0}\) is the college educated market potential for location \(c_0\). I then build up our estimating equation by subtracting the probability of staying in the home location resulting in

\[\ln \left(\frac{P(c|c_0)}{P(c_0|c_0)}\right) = \ln\left(\frac{T_{c_0c}}{T_{c_0c_0}}\right)+\theta\left((\omega_{c_0c}-\omega_{c_0c_0})-\lambda (p_{nc}-p_{nc_0}) - (p_{hc}-p_{hc_0})+(a_c-a_{c_0})\right)\]

I further make three simplifications to get to our key estimating equation. The first is to simplify the absolute advantage. I am currently allowing for there to be an absolute advantage in taste across specific origin-destination pairs. While this would help capture recurring flows as in \cite{Schubert2021}, limitations in the data make estimating these parameters challenging. Alternative approaches would be to allow for there to be an advantage for the origin location, but instead I will simply assume that there is no absolute advantage in taste for location after the education decision. Thus, I will make the simplification that \[T_{c_0c}= 1 \quad \forall \quad c_0,c \in C.\]

The second simplification is related to the effective wage. From Equation \ref{eq:effective_wage} and substituting in Equation \ref{eq:posterior_cov}, the effective wage for students with a college degree is 
\[\omega_{c_0c}= \mu_c^p-\frac{1}{2}\left(\sigma^2_{\xi}+  \frac{\left(\sigma^2_{\xi,c_0mc}+\sigma^2_{\hat{\xi},c_0mc}\right)}{\kappa M_{c_0c}\delta_{c_0}}\right)-t_{c_0}\]
Consider the migration matrix, \(M_{c_0c}\). Most individuals stay in their local area whereas a much smaller number migrates to a particular destination. This implies that the major impact on the signal strength is whether the student is considering the origin location vs. an alternative destination. Furthermore, I'll allow for the potential for an elasticity other than 1 with respect to the fraction of college workers in the origin. Together this implies

\[\omega_{c_0c} = \mu_c^p-\tilde{\sigma}^2_{\xi}- \tilde{\zeta}  \delta^{-\gamma}\left(1+\tau\ind_{c\neq c_0}\right)-t_{c_0}\]
where 
\[\tilde{\sigma}^2_{\xi}=\frac{1}{2}\sigma^2_{\xi} \quad \text{and} \quad \tilde{\zeta} = \frac12\frac{\left(\sigma^2_{\xi,c_0mc}+\sigma^2_{\hat{\xi},c_0mc}\right)}{\kappa } \]
and \(\tau\) represents the lower signal intensity associated with cities that are not the origin.

The last assumption is to set \(\theta =1\), which reduces the estimating equation to something that is more akin to a standard discrete choice model with Type 1 Extreme Value shocks. Altogether, we can substitute in the effective wage and make the stated simplifications to get the moment that we will be estimating in the data:

\begin{equation}
\label{eq:simplified_moment}
    \ln \left(\frac{P(c|c_0)}{P(c_0|c_0)}\right) =(w_c-w_{c_0})- \zeta  \delta^{-\gamma}-\lambda (p_{nc}-p_{nc_0}) - (p_{hc}-p_{hc_0})+(a_c-a_{c_0})
\end{equation}
where \(\zeta = \tilde{\zeta}\tau.\) On the right hand side of Equation \ref{eq:simplified_moment}, everything is known except for \(\zeta\) and \(\gamma\). This means we can rearrange the above equation to get our estimating equation. I further allow \(\zeta\) to vary by year as this will help soak up potential trends over time, so I estimate

\[\ln(Res) = \ln(\zeta_t)-\gamma \ln\delta \]
where \(Res = \ln\left(\frac{P(c_0|c_0)}{P(c|c_0)}\right)-(w_{c_0}-w_{c})+\lambda (p_{nc_0}-p_{nc})+ (p_{hc_0}-p_{hc})-(a_{c_0}-a_{c}).\) The results shown in Table \ref{tab:estimates} has the expected sign of \(\gamma\) and is statistically significant with standard errors clustered at the origin level. 

\begin{table}[ht]
\caption{Estimates of Key Model Parameters Based on Migration Probabilities}
\label{tab:estimates}
\centering
\begin{tabular}{rlrrrr}
  \hline
 & Estimate & s.d. & p-value \\ 
  \hline
\(\gamma\) & 0.61 & 0.09 & 0.00 \\ 
\(\zeta_{1980}\)  & 7.26 & 0.16 & 0.00 \\ 
 \(\zeta_{1990}\)  & 7.59 & 0.14 & 0.00 \\ 
 \(\zeta_{2000}\)  & 8.03 & 0.12  & 0.00 \\ 
   \hline
\end{tabular}
\end{table}
I also convert this estimate into the dollar impact comparing originating in a county with a twentieth percentile vs. eightieth percentile college fraction in Table \ref{tab:dollar_impact}. In columns (a) and (b) show the variation in college fraction across each census year investigated. Column (c) shows the impact of local college fraction through the informational channel in dollar terms. Shifting from a low college to high college origin increases the perceived value of college by \$1200 in 1980. We can then compare this to the standard deviation of the value across all locations which includes the impact of wages, prices and amenities. We can see that the impact of local labor markets through channels other than the wage can change the perceived utility by roughly 1/5 of the amount of all factors combined across locations.

\begin{table}[ht]
\caption{Impact of Local College Fraction in Dollar Terms}
\centering
\begin{tabular}{ccccc}
  \hline
    & (a) & (b)   & (c)  &(d)  \\ 
    Year & \(\delta_{P20}\) & \(\delta_{P80}\)  &  Impact & SD \\ 
  \hline
  \(1980\) & 0.16 & 0.26 &  \$1200 &  \hphantom{1}\$6300\\ 
  \(1990\)& 0.18 & 0.31 &  \$1500  & \hphantom{1}\$7100\\ 
  \(2000\) & 0.21 & 0.35 & \$2100& \$10900 \\ 
   \hline
\end{tabular}
\label{tab:dollar_impact}
\end{table}
\subsection{Agglomeration Elasticity}
The last value to estimate in order to perform our simple empirical exercise is the agglomeration elasticity. For this I again follow the simplification outlined in Section \ref{sec:comparing_to_lit}. I also simplify the origin-destination component such that \(h_{c_0c}=h_c\). In order to determine the elasticity, I simply perform an OLS regression of skilled wage on the fraction of skilled workers in each destination. 

\[\ln(w_c) = \ln(\rho_c h_{c})+\gamma_\alpha \ln\delta_c \]
The results are shown in Table \ref{tab:agglomeration_est}. Here we get a positive and significant value of the agglomeration coefficient. It is significantly smaller than the value obtained for the signaling elasticity. This implies a much larger impact of college fraction on wages when moving from the twentieth to the eightieth percentile as observed in Table \ref{tab:dollar_impact_agglomeration}.

\begin{table}[ht]
\caption{Estimating Agglomeration Effect of Destination College Fraction}
\centering
\begin{tabular}{rlrrrr}
  \hline
 & estimate & s.d.  & p-value\\ 
  \hline
 \(\gamma_\alpha\) & 0.22 & 0.06 & 0.00 \\ 
   \hline
\end{tabular}
\label{tab:agglomeration_est}
\caption*{\footnotesize This analysis takes the destination MSA across decadal census year as the unit of observation. This regression includes both MSA and Year fixed effects for the years 1980, 1990, 2000.}
\end{table}

\begin{table}[ht]
\caption{Impact of Destination College Fraction on Wage}
\centering
\begin{tabular}{ccccc}
  \hline
    & (a) & (b)   & (c)  &(d)  \\ 
    Year & \(\delta_{P20}\) & \(\delta_{P80}\)  &  Impact & SD \\ 
  \hline
  \(1980\) & 0.16 & 0.26 &  \hphantom{1}\$7900 &  \hphantom{1}\$6300\\ 
  \(1990\)& 0.18 & 0.31 &  \hphantom{1}\$9100  & \hphantom{1}\$7100\\ 
  \(2000\) & 0.21 & 0.35 & \$10300 & \$10900 \\ 
   \hline
\end{tabular}
\label{tab:dollar_impact_agglomeration}
\end{table}

\section{Experiment}
\label{sec:experiment}
In order to get a sense for how the local labor market affects student outcomes, I evaluate the partial equilibrium impact of a more equal distribution of college workers on student expected utility. This simplified model is not suited for a full structural analysis, but it can show us where some of the impacts of local labor markets are coming from and how they compare to agglomeration effects. The key parameter of interest is the market potential, \(\Phi\), which in equilibrium will be proportional to the expected indirect utility for both college and non-college students through Equation \ref{eq:college_choice_eq}. 

I consider what would happen if we distribute the total number of college workers across all MSAs such that \(\delta_c = \delta \quad \forall c \in C\). The key question is how does the market potential change in response to the change in college fraction across all locations. I can further break this down to include or exclude the component due to local labor market signaling. In our simplified model we have 

\[\Phi_{c_0}=\sum_{c \in C}\exp\left[\rho_c h_c \delta_c ^{\gamma_\alpha}-\tilde{\sigma}_\epsilon ^2-\tilde{\zeta}\delta_c^{-\gamma}(1+\tau_{c\neq c_0})-\lambda p_{n,c'} - p_{h,c'}+a_c'\right]\]
I will further simplify this by noting that the variance term \(\tilde{\sigma}_\epsilon ^2\) only adds a constant of proportionality. Further I will assume that the origin mean wage is known with certainty such that the only uncertainty is the wage in alternative destinations. Mathematically this implies that the signaling uncertainty reduces to \(\zeta\delta_c^{-\gamma}\). Lastly, the exponential causes issues in the partial equilibrium because it leads to explosive results. Therefore, our key parameter of interest will be an analogous term where I eliminate the exponential. This parameterization will allow me to explore the impact of the change in college fraction heuristically:

\begin{equation}
    \label{eq:utility_measure}
    \tilde{\Phi}_{c_0}=\sum_{c \in C}\left[\rho_c h_c \delta _c^{\gamma_\alpha}-\zeta_{c\neq c_0}\delta_c^{-\gamma}-\lambda p_{n,c'} - p_{h,c'}+a_c'\right]
\end{equation}

\begin{figure}[h]
\centering
\caption{Impact of Redistributing College Workers on Welfare}
\label{fig4}
\includegraphics[height = 3.5in]{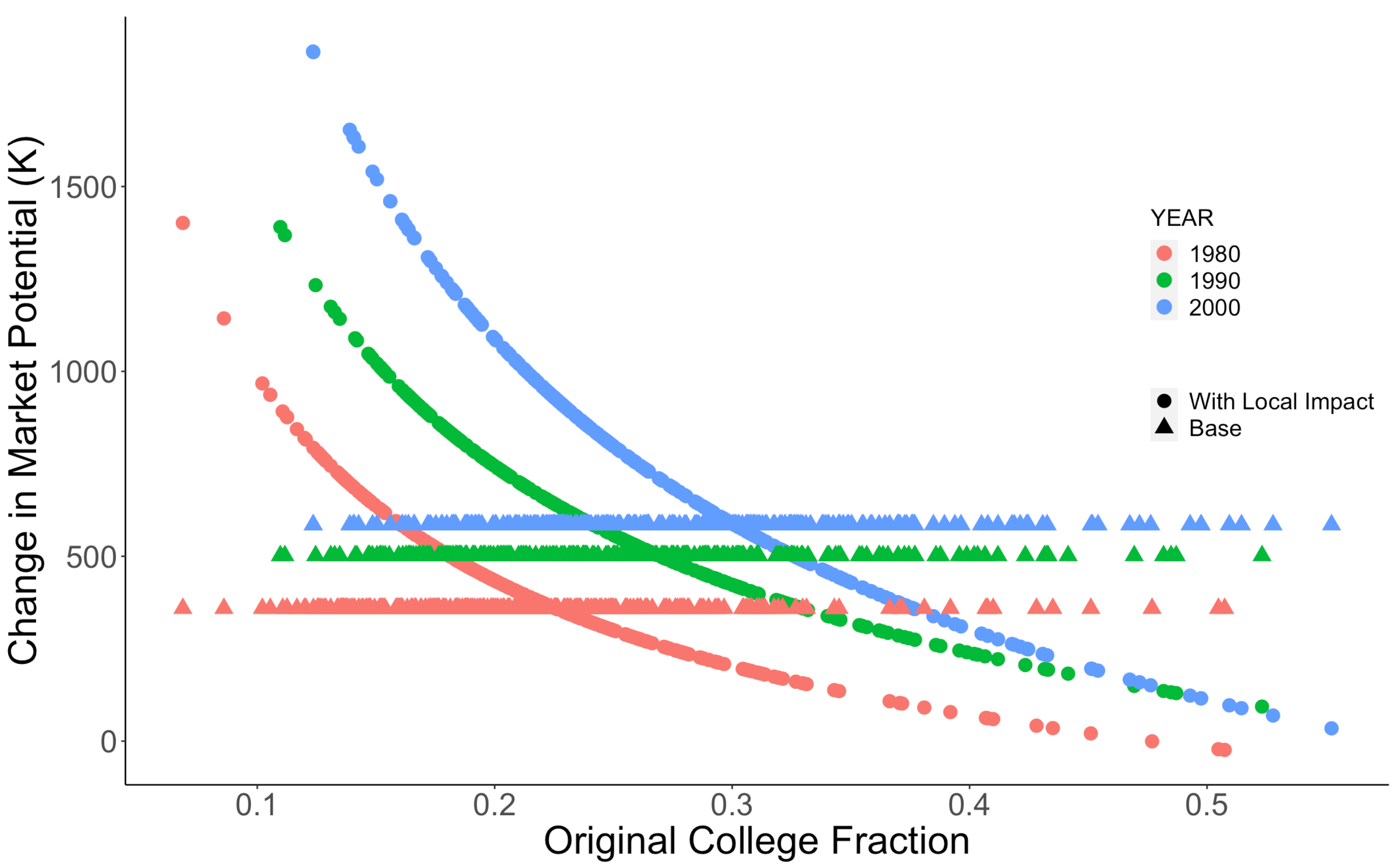}
\caption*{\footnotesize The change in a simplified measure of market potential are shown across different decades. This plot shows the change including the local labor market impact of signaling as well as without it. The baseline positive effect is due solely to the benefits of agglomeration from redistribution. }
\end{figure}

In Figure \ref{fig4}, I've plotted both the change in welfare associated with agglomeration alone as well as accounting for the local labor market signal. Because we have simplified the model to be a flat world without differential moving costs across destinations, the impact of the agglomeration component is the same across all locations independent of the original college fraction in that location. 

The effect of redistributing workers is positive here due to the concavity of the agglomeration function. The impact of skill redistribution across locations shows up in the impact on the risk term associated with signaling. Here we see that locations with low initial college fractions benefit substantially from redistributing skilled workers. These gains offset the losses experienced by students in initially high-skilled locations. Again, this is due to the convexity in the cost associated with a weak local labor market signal. These relative gains are also increasing across decades as the overall skill level in the population is increases. 

\section{Conclusion}
\label{sec:conclusion}
The main finding of this paper is that allowing for the composition of local labor markets to affect educational choice through signaling the returns to particular skills or similar channels can lead to significant heterogeneity in terms of the efficacy of place-based policies. This impact is an extension of the findings that there is substantial misinformation that students have with regards to the return to skill. The importance of signaling specifically aligns with evidence on the impact of role models and other interventions which can significantly affect major choice. I provide evidence in this paper that local labor markets tend to influence major choice beyond the affect on local wages. 

In light of these findings it is unsurprising that migration patterns have slowed as shown in Figure \ref{fig3}. Without major shake-ups in terms of labor market composition, we would expect the accumulation and concentration of particular skills in specific locations through a bias towards local labor market skills. This divergence would then lead to more mismatch between skills acquired in a particular location compared to those demanded in other locations. 

The objective of this paper is to spur further research into the spatial determinants of educational choice. Externalities which lead to biases in terms of skill acquisition related to local conditions (as is the case with signaling) will lead to spatial divergence in terms of outcomes and will reduce aggregate productivity. Policy recommendations which suggest the potential for concentrating/diversifying skills across space should makes sure to consider the impact that this can have on future generations of workers. Specifically, allowing for the spatial concentration of in-demand skills may substantially reduce the likelihood that students in less skill intensive regions invest in high return skills. It should be a policy goal to internalize these signaling externalities. 

These concerns should also be investigated to see how they interact with the availability of educational resources across space. Given that students tend to stay near home when pursuing their degrees the presence of educational deserts can be particularly devastating and add to the limited set of opportunities which students may consider. Beyond this, the resources at the major level may differ across space and may themselves be biased towards local labor markets. This may be advantageous when considering the creation of technology hubs but may make matching local skill acquisition to aggregate skill demand more challenging. 

A lot of attention has been focused on the evolution of the skill premium, but less research has been focused on how to ensure more people are able to benefit from the increasing skill premium. In this paper I propose an important explanation which has to do with the mismatch between local and global skill demands. However, there are many more reasons why we seem to fear the widening skill premium rather than embracing the ability for people to shoot up the ladder by taking advantage of the high wages associated with skill upgrading. I hope that more energy will be spent discussing this aspect of inequality.


\pagebreak 

\appendix
\section{Why Focus on Learning}
\label{sec:learning}

\subsection{Major Switching Between Freshman and Senior Year}

One way to test the informational channel is to consider what students learn once they enter college. As shown in \cite{NBERw28353}, students update their beliefs throughout college, developing more accurate predictions. Given the evidence that expectations of future earnings affect a student's choice of major, it is unsurprising that many students switch their majors after acquiring new information. Looking at the College Senior Survey performed by HERI which can be linked to the True Freshman Survey, I observe that roughly 40\% of students switch majors. This is roughly in line with data from the National Center for Education Statistics. We might expect students with more information about wages for their stated major to be less likely to switch majors as they are less likely to experience a negative shock to their expectations.

In line with the hypothesis that local labor markets provide information about potential majors, table \ref{tab:maj_switching} shows that students that state their intentions to major in a field which is more highly represented in their local labor market have a significantly lower likelihood of switching out of this degree. This is suggestive that students may be updating less about their intended major when it is well represented in their local labor market. The effect remains significant with a similar point estimate after controlling for major or state fixed effects. 

\begin{table}[H]
\caption{Effect of Labor Market Similarities on Major Switching \label{tab:maj_switching}}

\global\long\def\sym#1{\ifmmode^{#1}\else$^{#1}$\fi}%
 \begin{center}
\begin{tabular}{llll}

\hline 
 & \multicolumn{3}{c}{Fraction Switching}\tabularnewline
\hline 
Fraction of Workers with Major in County & -0.87\sym{{***}} & -0.60\sym{{***}} & -0.53\sym{{**}} \tabularnewline
 & (0.09) & (0.18) & (0.18)\tabularnewline
\hline 
Major FE & NO & YES & YES\tabularnewline
State FE & NO & NO & YES\tabularnewline
\hline 
Obs & 68142 & 68142 & 68142\tabularnewline
$R^{2}$ & 1.5\% & 4.2\% & 4.4\%\tabularnewline
\hline 
\multicolumn{3}{l}{{\tiny{}Robust standard errors clustered at the primary industry in
parentheses.}}\tabularnewline
\multicolumn{3}{l}{{\tiny{}\sym{*}$p<0.05$, \sym{**}$p<0.01$,\sym{***}$p<0.001$.}}\tabularnewline
\end{tabular}
\end{center}
\end{table}

\subsection{Information Flows Through Local Major and Industry Composition}

In an attempt to distinguish the underlying mechanism behind the informational signals from local labor markets, I compare the importance of the local educational, occupational and industrial composition on migration flows. Counties which are more similar in terms of these factors might be expected to have more migration between them. This may be due to companies relocating workers across cities or because skills learned on the job are more transferable across cities. However, it might also be the case that informational flows through these channels such that workers feel more confident migrating to areas with similar labor market structures. For instance, someone in Boston which has a relatively high density of software engineers is more likely to know the wage of software engineers in San Francisco. This is likely true across occupations, industries and degrees. 

As a test to see whether this is true in the data I form a simple measure of proximity in terms of occupation, industry and education. Define matrix \(O\), such that each column represents the occupation distribution in a particular county, \(c\). Therefore \(O_{co}\) is the fraction of prime aged workers in county \(c\) with occupation \(o\). The likelihood that someone in that occupation receives information about a different occupation is proxied by the occupational transition matrix between the two which is estimated from the CPS. Thus, for occupation we define the proximity in terms of occupation between \(c\) and \(c'\) as 

\[\Sigma^o_{cc'} = O_cTO_{c'}\]

Additionally, the total level of migration is going to depend on the population in \(c\), \(L_c\), and in \(c'\), \(L_{c'}\), as well as the distance between locations. Using Census data for the year 2000, I estimate the following regression: 

\[\log(M(c, c')) = \alpha +  \beta \log(\Sigma) +  \beta_2 \log( L_c)  + \beta_3 \log( L_c)' + \eta \log(\delta(c,c'))\]
where \(M(c, c')\) is the number of movers and  \(\delta(c,c')\) is the distance between \(c\) and \(c'\). The results of this regression focusing on movers between 25 and 35 years old are shown in table \ref{tab:local_mig}. I focus on this age range as it is following graduation and is most likely to be a move that could have been anticipated when making educational decisions. Column 4 shows that when taken together industrial and educational similarities are more important determinants of migration patterns for young movers. When I limit the sample to movers who are older than 35 years of age, the impact of educational similarity disappears whereas occupational and industrial composition becomes more significant. This is what you would predict if skills acquired overtime in a particular occupation or industry become more important determinants of the ability to find a new job elsewhere. However, the evidence for young workers where these skills haven't been built up are more likely to be due to information especially in light of the information above that suggests that along with developing skills in college, students are improving their estimate of expected earnings and that this extends to expected earnings across locations.

\begin{table}[H]
\caption{Fraction of Students Switching Majors \label{tab:local_mig}}
\global\long\def\sym#1{\ifmmode^{#1}\else$^{#1}$\fi}%
\begin{center}
\begin{tabular}{llllll}

\hline 
 & \multicolumn{4}{c}{  Movers with \(25\leq\) Age \(\leq 35\)} & Age \(> 35\) \tabularnewline
\hline 
Occupational Similarity & 0.36\sym{{**}} & &  &0.21\sym{{.}} & 0.14\sym{{***}}\tabularnewline
 & (0.12) & & &(0.12) & (0.03)\tabularnewline
 Industrial Similarity &  & 0.31\sym{{***}} & & 0.25\sym{{**}}  & 0.14\sym{{***}}\tabularnewline
 & & (0.07) & &(0.08)&(0.02) \tabularnewline
 Educational Similarity & & & 0.18\sym{{**}} & 0.13\sym{{*}}  & -0.00\tabularnewline
 &  & & (0.06)&(0.06)& (0.01)\tabularnewline
\hline 

\hline 
Obs & 16583 & 16583 & 16583 & 16583\tabularnewline
$R^{2}$ & 1.5\% & 4.2\% & 4.4\% &\tabularnewline
\hline 
\multicolumn{5}{l}{{\tiny{}Robust standard errors clustered at the primary industry in
parentheses.}}\tabularnewline
\multicolumn{5}{l}{{\tiny{}\sym{*}$p<0.05$, \sym{**}$p<0.01$,\sym{***}$p<0.001$.}}\tabularnewline
\end{tabular}
\end{center}
\end{table}

\bibliography{refs.bib}

\pagebreak

\end{document}